\documentclass[12pt,pre,superscriptaddress,showpacs]{revtex4}
\usepackage{graphicx}
\usepackage{dcolumn}
\usepackage{bm}

\begin{document}

\title
{Molecular dynamics simulations of crystallization of hard
spheres}

\author{Igor Volkov}
\affiliation{Department of Physics, 104 Davey Laboratory, The
Pennsylvania State University, University Park, Pennsylvania
16802}

\author{Marek Cieplak}
\affiliation{Department of Physics, 104 Davey Laboratory, The
Pennsylvania State University, University Park, Pennsylvania
16802}

\affiliation{Institute of Physics, Polish Academy of Sciences,
02-668 Warsaw, Poland}

\author{Joel Koplik}
\affiliation{Benjamin Levich Institute and Department of Physics,
City College of the City University of New York, New York, NY
10031}

\author{Jayanth R. Banavar}
\affiliation{Department of Physics, 104 Davey Laboratory, The
Pennsylvania State University, University Park, Pennsylvania
16802}

\date{\today}

\begin{abstract}
We have carried out  molecular dynamics simulations of the
crystallization of hard spheres modelling colloidal systems that
are studied in conventional and space-based experiments. We use
microscopic probes to investigate the effects of gravitational
forces, polydispersity and of bounding walls on the phase
structure. The simulations employed an extensive exclusive
particle grid method and the type and degree of crystalline order
was studied in two independent ways: by the structure factor, as
in experiments, and through local rotational invariants. We
present quantitative comparisons of the nucleation rates of
monodisperse and polydisperse hard sphere systems and benchmark
them against experimental results. We show how the presence of
bounding walls leads to wall-induced nucleation and rapid
crystallization and discuss the role of gravity on the dynamics of
crystallization.
\end{abstract}


\maketitle

\section{\label{sec:intro}Introduction}

Hard sphere systems are idealized approximations to a large number
of physical systems, such as simple liquids \cite{1}, glasses
\cite{2}, colloidal dispersions \cite{3} and particulate
composites \cite{4} and are now being studied extensively in a
microgravity environment \cite{7, 4-1, weitz} which allows for a
creation of new technological materials, such as photonic crystals
\cite{phot}. The use of colloidal particles for engineering new
materials  is a relatively unexplored field which promises to
revolutionize materials synthesis. Colloidal suspensions are also
interesting from a fundamental scientific point of view since they
self-assemble into a wide range of structures.  Thus, they  may be
thought of as models of atomistic condensed matter systems with
the distinct advantage of relevant length and time scales being
more readily accessible to experiments.

On Earth, the effects of sedimentation and gravity-induced
convection can cloud, modify, or sometimes even radically alter,
the intrinsic behavior of certain classes of colloidal systems.
Because the binding energies of the crystalline phases are low and
comparable to each other, gravity can greatly influence the
kinetics of formation and, indeed, the very nature of the observed
crystal structure. Colloidal suspensions of hard spheres are model
systems for studying the statistical mechanics of structural phase
transitions. Such suspensions undergo an entropy-driven phase
transition from fluid to crystal as a function of increasing
volume fraction. Unlike comparable phase transitions in
conventional systems of condensed matter, the dynamics of such
structural phase transitions can be monitored with ``atomic''
precision using conventional light microscopy. In hard sphere
systems, at high volume fractions, glass formation competes with
the nucleation and growth of the crystalline phase. The
Chaikin-Russel experiments on the Space Shuttle \cite{7,4-1} have
led to the striking result that  samples of hard sphere colloids
that remain glassy on the Earth for more than a year crystallize
within a few weeks in a microgravity environment.

In this paper, we present results of molecular dynamics (MD)
simulations of the crystallization of hard spheres. These
simulations allow for microscopic probes of the physics involved
in both conventional and space-based measurements of nucleation
and crystal growth in colloidal systems. We focus on the effects
of weak gravitational forces, polydispersity and on the effects of
bounding walls on phase structure. We present quantitative
comparisons of the nucleation rates of monodisperse and
polydisperse hard sphere systems and benchmark them against
experimental results. We demonstrate that the presence of gravity
can delay crystallization. Furthermore, we show how the presence
of the bounding walls leads to wall-induced nucleation and rapid
crystallization.

\begin{figure}
\includegraphics[scale = 0.7]{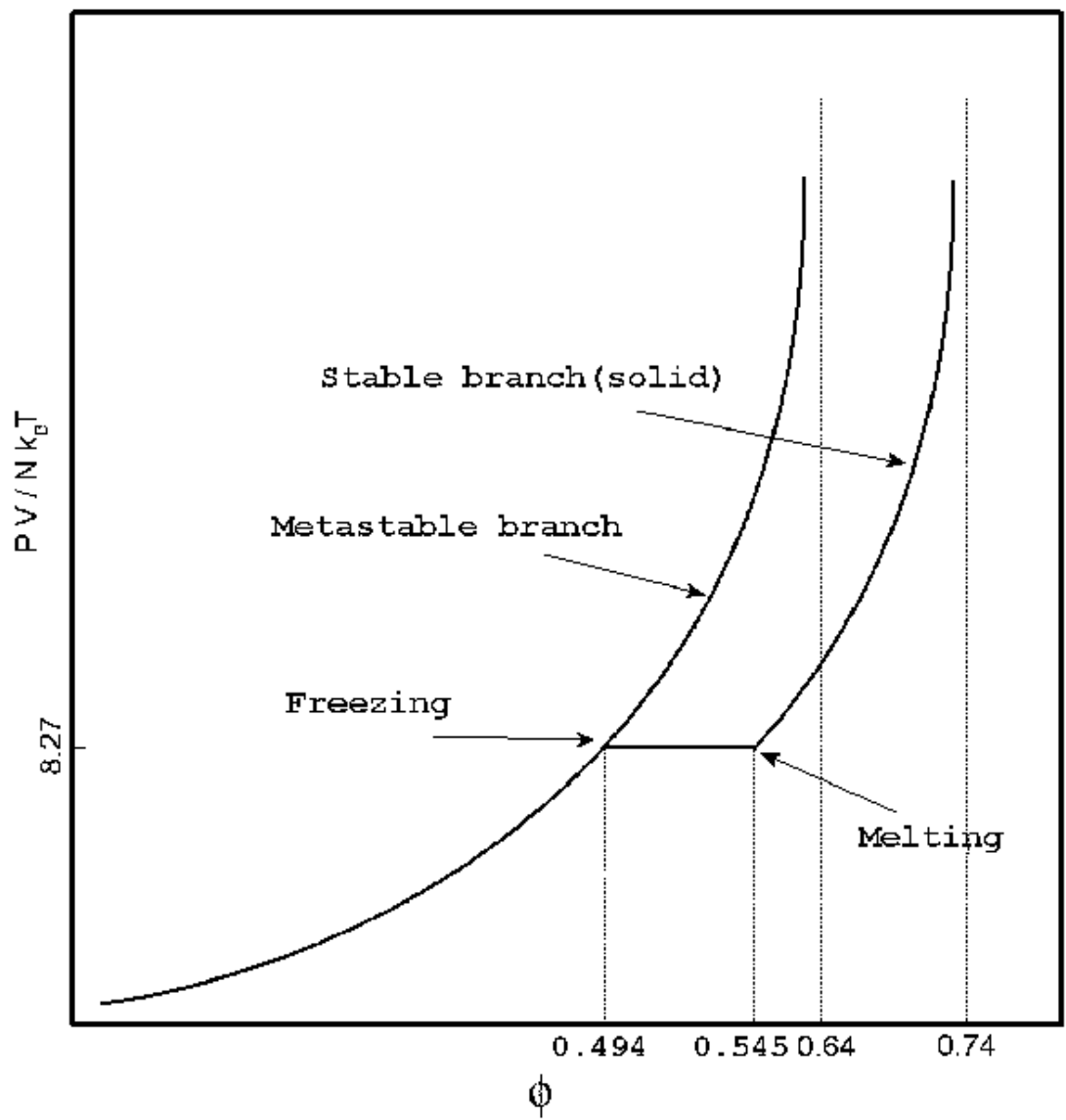}
\caption{\label{fig:phase} Schematic representation of the phase
diagram of a system of hard spheres. }
\end{figure}

Numerical studies of the hard sphere system started with the
pioneering work of Alder and Wainwright \cite{5}. Since then,
there have been many studies that elucidated the nature of the
phase diagram. In particular, computer simulations(see
\cite{6,n1,n2,6-1} for a few examples) have provided evidence for
the existence of a first order fluid-to-solid transition in the
hard sphere system. With an increase of the packing fraction,
$\phi$, (defined as the ratio of the volume occupied by the
spheres to the total volume) the system in the liquid state
reaches the freezing point at $\phi = 49.4\%$ (see
Fig.~\ref{fig:phase} for a sketch). The phase diagram splits into
metastable and stable branches at this point. The metastable
branch is a continuation of the liquid branch and it exists in the
region between the freezing point and $\phi \approx 64\%$ which
corresponds to the random close packing (rcp) state. The rcp
provides the maximum $\phi$ that can be achieved in the disordered
system. The stable branch consists of a coexistence region of
liquid and crystal which ends at $\phi$ of $54.5\%$ corresponding
to the melting point. Above the melting point, the stable branch
represents the crystal state and that is present up to
$\phi\approx74\%$ which corresponds either to the close packed
face-centered cubic (fcc) or to the hexagonal closed packed (hcp)
configurations.

The metastable branch, especially its part above the melting
point, has received a lot of attention  in the last several years.
One of the debated issues here is the existence of the glassy
state in the metastable system when $\phi > 58\%$, i.e. in the
vicinity of the rcp value. A number of papers report no sign of
crystallization \cite{6-1,6-2,6-3} and thus confirm the presence
of the glassy state. On the other hand, Rintoul and Torquato
\cite{6} have argued that if computer simulations were to run for
a sufficiently long time, then crystallization would eventually
set in. A striking experimental evidence for this scenario has
been provided by a recent microgravity experiment on the Space
Shuttle \cite{7}. It demonstrated crystallization  in a hard
sphere colloidal dispersion at $\phi = 61.9\%$ occurring on the
time scale of several days whereas the same system stayed
amorphous for more than a year when studied on Earth.

The formation of the crystals in the supersaturated hard-sphere
system is commonly described by the classical nucleation theory
(see \cite{n5} and references therein). According to this theory,
a crystallite forms in the system due to thermal fluctuations and
then its total free energy consists of two terms: a negative bulk
term, which is proportional to the volume of the crystallite, and
a positive surface term which is proportional to its surface area.
This leads to the prediction that the crystallite will continue to
grow only when its size is bigger than a certain critical value
and it will shrink otherwise. There are a number of experimental
results that support the classical nucleation theory \cite{n5,17}.

The MD simulations of the hard spheres systems that we report on
in this paper are focused on the dynamics of crystallization above
the melting concentration and are complementary to the earth-based
studies of Gasser et al. \cite{gasser1}. The crystallization
process is monitored by means of local order parameters as well as
through the static structure factor. The former method is
currently widely used to analyze the results of computer
simulations whereas the structure factor is measured
experimentally. We investigate the influence of bounding walls,
polydispersity and of gravitational field on the dynamics of
crystallization and show that the nucleation rates for
crystallization are comparable to the values obtained
experimentally.

We show that the system with periodic boundary conditions
crystallizes in a somewhat complex manner with an interconnected
phase of growing crystal nuclei. In contrast, a system with planar
walls exhibits layering and leads to a heterogeneous wall
nucleation mechanism characterized by more rapid crystallization.
For volume fraction around $56\%$, gravity leads to a
concentration gradient accompanied by the formation of very
well-defined layers with excellent planar ordering. However, at
larger volume fractions, gravity causes the crystallization
process to slow down relative to the planar wall case without any
imposed gravitational field. Polydispersity in the size
distribution of the hard spheres leads to slower crystallization
and in the absence of gravity, we found an increase with time of
the relative fraction of hard spheres with fcc order compared to
hcp suggesting that the former crystal structure is preferred to
the latter.

The outline of the paper is as follows. In Section~\ref{sec:MD},
we describe the algorithms used in the simulations. In
Section~\ref{sec:char}, we present the methods of the analysis of
the local structure and of the thermodynamical properties of the
system. Section~\ref{sec:dynamics} presents the results of our
simulations for both monodisperse and polydisperse systems with
periodic boundary conditions. Section~\ref{sec:gravity} considers
the effects arising due to rigid flat walls that restrict motion
in one direction and discusses the role of a uniform gravitational
field along this direction. Finally, in
Section~\ref{sec:structure}, we discuss the nature of the
crystalline phase.

\section{\label{sec:MD}The MD simulation}

There are many possible algorithms that can be used in the MD
simulations of hard sphere systems \cite{10}. Owing to the
simplicity of the potential, the only events that need to be
calculated are the consecutive collisions between the particles.
In this respect, the MD algorithms for the hard sphere systems are
quite distinct from the algorithms for the soft types of
potentials where the evolution between the collisions also
matters. Thus the evolution should not be considered in equal time
steps but instead it ought to be studied through an event driven
algorithm. The most challenging part of such an algorithm, in
terms of its computational performance, is the proper scheduling
of the future collisions and the organization of the data
structure.

Our MD simulations were performed by implementing the algorithms
proposed by Isobe \cite{11} who introduced the concept of an
extended exclusive particle grid method to the studies of hard
sphere and hard disk systems. In this method, the volume $V$
containing the particles is divided into small cells, so that each
cell contains no more than one particle. Thus, the continuous
coordinates of the particles are ``mapped'' onto a lattice which
allows for an easy specification of neighboring particles.
Candidates for the next particle-pair collision are found just by
searching the neighboring cells. Once this is accomplished, the
next collision event for the system can be found by creating a
complete binary tree \cite{12}. The positions of all the particles
do not need to be updated after each collision, since in a
sufficiently dense system the neighborhood of a particle remains
the same for a long time.

The initial packing of the system of $N$ hard spheres was
generated from a random set of points within a box by using an
iterative algorithm proposed by Jullien et al. \cite{13}. At each
stage of this algorithm one identifies the pair of particles with
the smallest mutual distance $d_m^i$ (the superscript $i$ refers
to the $i$th stage of the iterative procedure) and moves them
apart symmetrically by a distance $d_M^i$ which decreases with
each iterative step according to the following formula:
\begin{equation}
d_M^{i+1} = d_M^i - {\hat{R} \over N} \left(\phi_M^i - \phi_m^i
\right)^{1/3} \;\;.
\end{equation}
Here $\phi_{M,m}^i = \pi d_{M,m}^3 N / (6V)$, $\phi_M^0=1$, and
$\hat{R}$ is a parameter of the algorithm. The process continues
until $d_M < d_m$ and the final value of $d_m$ is chosen to be the
particle diameter. Different values of $\hat{R}$ lead to different
packing fractions and generally, the smaller the $\hat{R}$, the
larger the packing fraction. In the limit of $\hat{R}\rightarrow
0$, one reaches a packing fraction corresponding to the random
close packed value. In order to obtain a polydisperse distribution
of the radii we modified this algorithm so that at each iteration
step we move apart two particles that overlap the most and their
new mutual distance is set equal to the sum of the predefined
particles' radii.

Our MD simulations were performed with at least $10976$ hard
spheres (both in the mono- and polydisperse cases). The particles
 were placed in a cubic box. In the absence of any
walls, periodic boundary conditions were imposed. When studying
the effects of the walls, two flat walls were introduced at $z=0$
and $z=L$ while maintaining the periodic boundary conditions in
the other two directions. This was accomplished by changing the
standard algorithm \cite{13} so that the walls are represented by
two new ``particles'' which do not move. The initial particle
velocities were chosen to be random with a Gaussian distribution
and zero total momentum.

The results were averaged over six simulations for each set of
control parameters. We have focused on the concentration range
from $\phi =54 \%$ to $\phi = 58 \%$ for systems without the
bounding walls and gravity and from $\phi = 54 \%$ to $\phi = 63
\%$ in the other cases. This procedure was motivated by the fact
that for lower and higher concentrations the crystallization times
increase substantially and so does the computational time.

In our simulations we define the hard sphere diameter to be $1$
unit and the timescale is defined by choosing the the mean
absolute velocity of the hard spheres to be $1$. Following the
approach of Harland and van Megen \cite{17}, in order to make
contact with experiment, we show the results of our simulations by
expressing times and lengths in units of the diffusional
characteristic time $\tau_b = R^2 / D_0$ and hard sphere diameter
$2 R$, respectively. Here, $\displaystyle{D_0 = {3\pi\over
16\sqrt{2}} \bar{v} l_{mfp}}$, where $\bar{v}$ is the mean
absolute velocity of the hard spheres and the mean free path
$\displaystyle{l_{mfp} = {V\over N 4\pi R^2}}$. The acceleration
due to the gravity was chosen to be approximately $4.7$ (see
caption in Figure~\ref{fig:cp} for precise values) in units in
which the hard sphere diameter is $1$ and the mean absolute
velocity is $1$.

\section{\label{sec:char}Characterization of the hard sphere systems}

\subsection{The equation of state}

The relevant parameter that describes the thermodynamic properties
of the hard sphere system is the pressure, $P$, since the internal
energy of such a system is that of an ideal gas. Changing the
temperature, $T$, is simply equivalent to rescaling the time
scale. The pressure can be calculated by using the radial
distribution function or through the collision rate in the system.
The latter method is more reliable because of the difficulties
with a precise determination of the radial distribution function.

The equation of state in terms of the collision rate $\Gamma$ is
given by \cite{14}
\begin {equation}
{P V\over N k_B T} = 1 + {\Gamma\over\Gamma_0}{B_2\over V}\;\;,
\end{equation}
where $V$ is the total volume, $N$ the number of particles, $k_B$
the Boltzmann's constant, $B_2$ the second virial coefficient.
$\Gamma_0$ is the low-density collision rate which is given by
\cite{14-1}
\begin{equation}
\Gamma_0 = 8 {N (N - 1) \over V}R^2 \sqrt{\pi < v^2 > \over 3}
\;\;,
\end{equation}
where $< v^2 >$ is the mean square velocity and $R$ the radius of
the sphere.

The pressure was monitored throughout the simulation and was used
as a quantitative parameter which allowed us to check on the
progress of the crystallization.

\subsection{The local structure}

A number of methods has been used in the literature to
characterize the local structure and a degree to which it is
crystalline. A widely used technique to distinguish between
crystalline and amorphous structures is through the Voronoi
analysis of the topology of the neighborhood of a given particle.
The Voronoi polyhedron is defined \cite{2} as the set of all
points that are closer to a given particle than to any other.
Partitioning of space into the Voronoi polyhedra allows one to
make a natural identification of the neighbors. Determination of
the numbers of walls in the Voronoi polyhedra leads to an
unambiguous selection of the particles in the solid-like regions.
However, such an analysis lacks precision when applied to
thermally distorted crystals and is not too effective in
distinguishing between various types of crystalline order. The
same difficulties arise when the structure, crystalline or not, is
analyzed through the particle distribution function.

\subsubsection{The local invariants}

In order to determine the kind of the local order around a
particle and to distinguish between the fcc, hcp, bcc and
liquid-like configurations we make use of the local order
parameter method \cite{15,16}, which gives reliable results even
in the case of crystalline structures which are highly perturbed.
The first step here is to construct the normalized order parameter
$ \hat{q}_{lm}$ for a particle $i$ through
\begin{equation}
  \hat{q}_{lm}(i)={1\over
  N_b(i)}\sum_{j=1}^{N_b(i)}Y_{lm}(\vec{r}_{ij}) \;\;,
\end{equation}
where $N_b(i)$ is the number of neighbors of the particle,
$Y_{lm}$ is a spherical harmonic and $\vec{r}_{ij} =
\vec{r}_j-\vec{r}_i$ with $\vec{r}_{i}$ being the coordinates of
the center of particle $i$. The neighbors are defined to be those
particles which have a mutual distance less than a certain cutoff
value. It is physically appealing to choose the cutoff as
corresponding to the position of the first minimum in the radial
distribution function. $Y_{lm}$ is the spherical harmonic function
which means that $\hat{q}_{lm}(i)$ has $2l+1$ complex components.
$\hat{q}_{lm}(i)$ can be normalized by multiplication of a
suitable constant to yield $\bar{q}_{lm}(i)$, such that
\begin{equation}
  \sum_{m=-l}^{m=l}\bar{q}_{lm}(i)\bar{q}_{lm}^*(i)=1\;\;.
\end{equation}
If
\begin{equation}
\label{eqn6}
\left|\sum_{m=-l}^{m=l}\bar{q}_{lm}(i)\bar{q}_{lm}^*(j)\right|>0.5
\end{equation}
then the bond between particles $i$ and $j$ is considered to be
crystal-like. Furthermore, if a particle has seven or more
crystal-like bonds, then it is counted as belonging to a
crystalline region. Note, that $\bar{q}_{lm}(i)$ is not
rotationally invariant and hence the quantity on the left hand
side of Eq.~(\ref{eqn6}) depends on the choice of the coordinate
axes. Indeed, for a given bond, there can be ambiguity about
whether the quantity in Eq.~(\ref{eqn6}) is greater than the
threshold value of $0.5$ or not. However, when summing over all
the bonds connected to a given hard sphere, the criterion for
crystallinity is substantially independent of the choice of the
coordinate axes.

In order to distinguish between different crystal structures we
construct the second-order rotational invariants $q_4(i)$,
$q_6(i)$, and $\hat{w}_6(i)$ \cite{16-1}, where
\begin{equation}
  q_l(i)=\left[{4\pi\over
  2l+1}\sum_{m=-l}^{m=l}|\hat{q}_{lm}(i)|^2\right]^{1\over2}\;\;
\end{equation}
and
\begin{equation}
  \hat{w}_l(i)=
  \mathop{\sum_{m_1,m_2,m_3}}_{m_1+m_2+m_3=0}
  \left(
  \begin{array}{ccc}
    l  &  l  &  l  \\
   m_1 & m_2 & m_3 \\
  \end{array}
  \right)\hat{q}_{lm_1}(i)\hat{q}_{lm_2}(i)\hat{q}_{lm_3}(i),
\end{equation}
where $\left(
  \begin{array}{ccc}
    l  &  l  &  l  \\
   m_1 & m_2 & m_3 \\
  \end{array}
  \right)$ is a Wigner $3j$ symbol \cite{wigner}. After
calculating $q_4(i)$, $q_6(i)$, $\hat{w}_6(i)$ one can decompose a
vector $\vec{s}$ consisting of these three components into the
five characteristic vectors $\vec{s}_{fcc}$, $\vec{s}_{hcp}$,
$\vec{s}_{bcc}$, $\vec{s}_{sc}$ and $\vec{s}_{ico}$ corresponding
to perfect fcc, hcp, bcc, sc, and icosahedral structures. The
values for the perfect crystals are given in
Table~\ref{tab:table1}. Such a decomposition can be carried out by
minimizing the following expression \cite{tw95}:

\begin{equation}
\Delta^2 = \left[\vec{s} - (f_{fcc}\vec{s}_{fcc} +
f_{hcp}\vec{s}_{hcp} +f_{bcc}\vec{s}_{bcc} +f_{sc}\vec{s}_{sc}
+f_{ico}\vec{s}_{ico})\right]^2
\end{equation}
with a constraint that all of the $f$ factors are positive and
they add up to 1. As a result we get a set of five numbers $f$.
Each $f$ represents the ``importance'' of the corresponding
structure. For example, for each particle of the perfect fcc
crystal we would get $f_{fcc}=1$ and all the others to be zero.
For an imperfect crystal, we assign each particle to the structure
corresponding to the biggest $f$. Note, that our method is
slightly different from that used in Ref.~\cite{tw95} but in
practice the two methods yield similar results. In
Ref.~\cite{tw95}, the clusters of particles were analyzed by
comparing the distributions of the local order parameters for a
given cluster and thermally equilibrated perfect crystals.

\section{\label{sec:dynamics}Dynamics of crystallization of mono- and
polydisperse systems}

We begin with an analysis of the crystallization process as
monitored through the evolution of the Bragg peak in the static
structure factor $S(q)$ \cite{17} where $q$ is the wave number.
This method is widely used in analyzing data in the light
scattering experiments.

After isolating the Bragg peak in the structure factor curve, we
remove the liquid contribution by subtracting the Percus-Yevick
result \cite{18} multiplied by a constant which varies from $0$
(in the fully crystallized state) to 1 (in the liquid state) in
order to ensure that $S(q)\rightarrow 0$ at small $q$. The crystal
fraction, $X$, can be found by integrating the Bragg peak and
choosing the upper limit of the integration at the minimum of
$S(q)$ and by normalizing the result, so that $X=1$ in the fully
crystallized state. The other parameters which can be determined
in this approach are the average linear crystal size, $L = 2 K \pi
/ \Delta q$, where $K=1.155$ is the Scherrer constant for a
crystal of a cubic shape \cite{sch}, the number density of the
crystals, $N_c = X / L^3$, and the nucleation density rate, $I = d
N_c / dt$ \cite{17}.

\begin{figure}
\includegraphics[scale = 0.7]{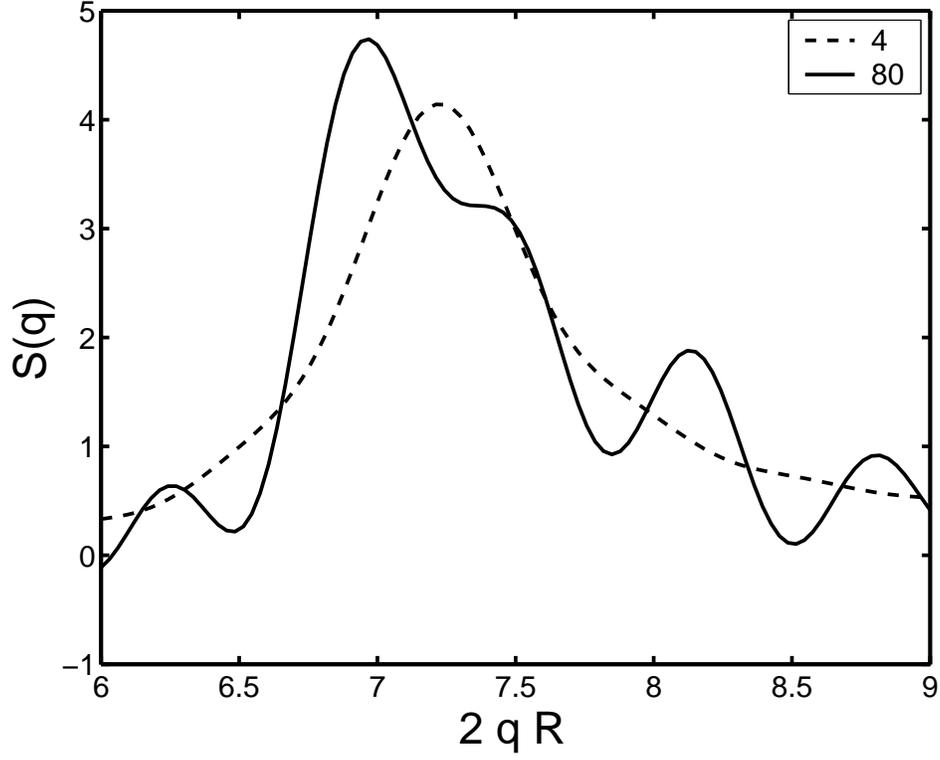}
\caption{\label{fig:sf55} The dependence of the static structure
factor on the wave number for $\phi = 55 \%$. The two curves shown
correspond to the different stages of crystallization (after 4 and
80 steps where one step counts as 500 $\times N$ collisions, where
$N$ is the number of particles). The dashed-line curve represents
the Percus-Yevick solution (the liquid state, 4 steps) and the
solid curve represents the fully crystallized system (80 steps). }
\end{figure}

\begin{figure}
\includegraphics[scale = 0.7]{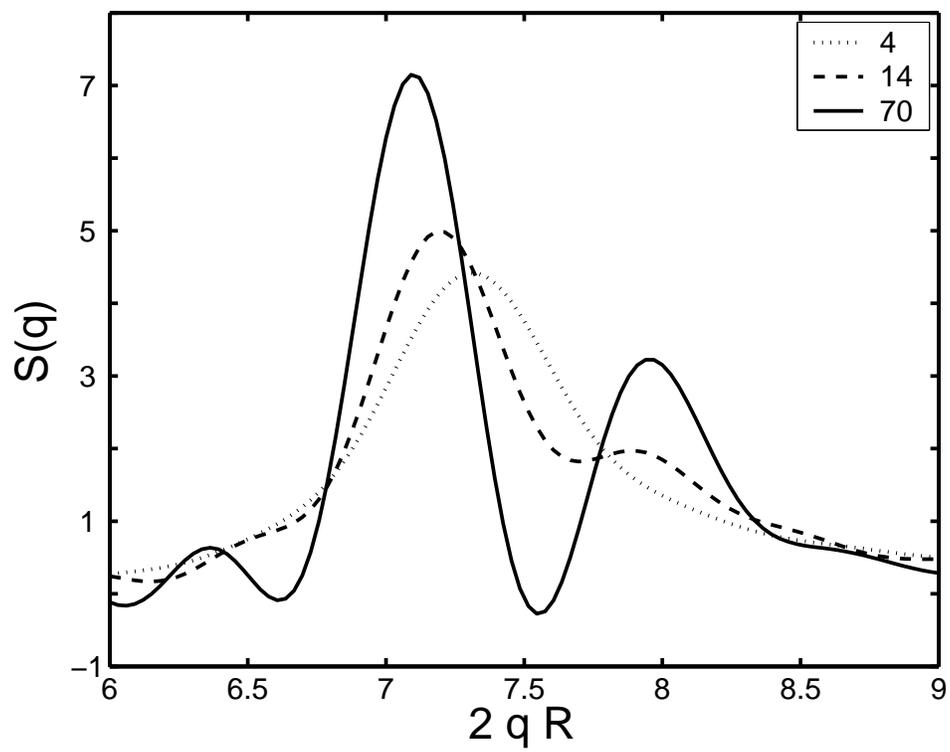}
\caption{\label{fig:sf58} The dependence of the static structure
factor on the wave number for  $\phi = 58 \%$. The curves shown
correspond to the three different stages of crystallization. The
step numbers are indicated at the right hand corner. }
\end{figure}

An example of the time variation of the static structure factor
for the monodisperse system (at $\phi = 55 \%$) is shown in
Fig.~\ref{fig:sf55}. One observes that the structure factor
exhibits the expected dynamics, namely, the Bragg peak at $2 q R
\approx 7$ corresponding to the $\{111\}$ direction becomes higher
and higher and it shifts to lower wave numbers on crystallization.
However, it is difficult to isolate the Bragg peak due to the
emergence of other peaks, for instance, of the one corresponding
to the fcc structure ($\{200\}$ peak). Note, that the shape of the
structure factor on the left hand side of the Bragg peak remains
substantially unchanged. Therefore, for the analysis of the
structure, we used only the left half of the Bragg peak and then
multiplied the results by a factor of two. For example, in
Fig.~\ref{fig:sf55}, the lower integration limit was taken to be
6.5 and the upper one at the maximum of the Bragg peak. At higher
packing fractions, we observed distinctive Bragg peaks at all
stages of crystallization (Fig.~\ref{fig:sf58}).

\begin{figure}
\includegraphics[scale = 0.7]{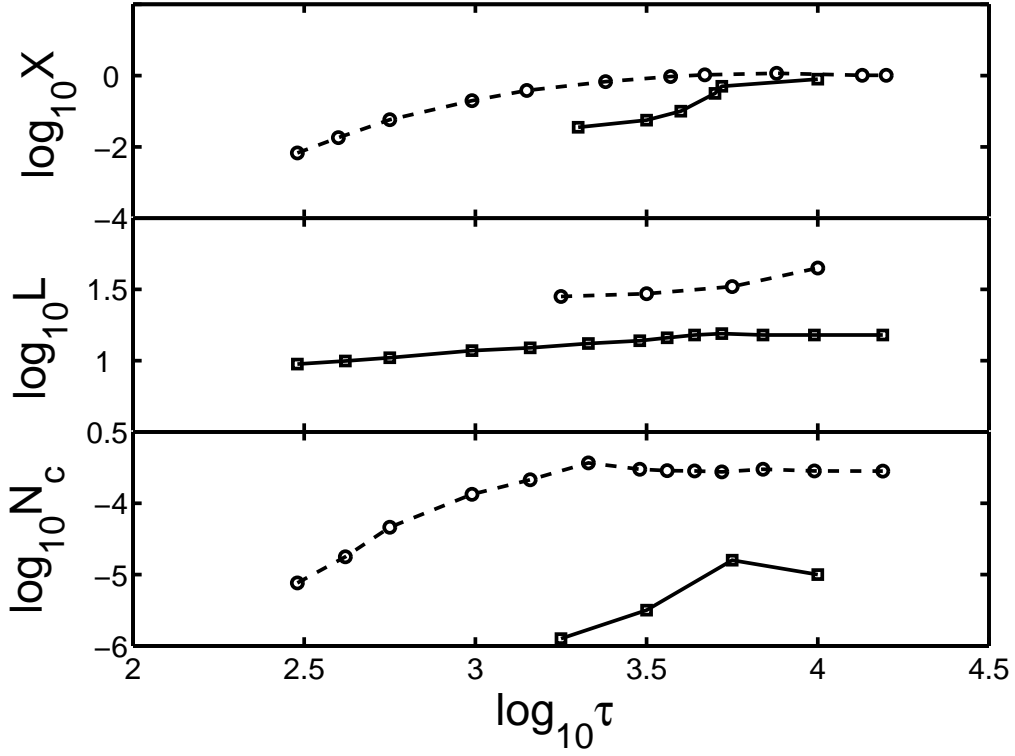}
\caption{\label{fig:xlnc} The time evolution of the
 crystal fraction $X$(top panel), the average linear
crystal size $L$ (in particle diameters, middle panel), and the
number density of the crystals $N_c$ (in units of $(2 R)^{-3}$,
bottom panel) for the monodisperse system at $\phi = 55 \%$. Our
results are shown as solid lines and the experimental results
\cite{17} as dashed lines. The time is measured in units of the
diffusional characteristic time. }
\end{figure}

\begin{figure}
\includegraphics[scale = 0.7]{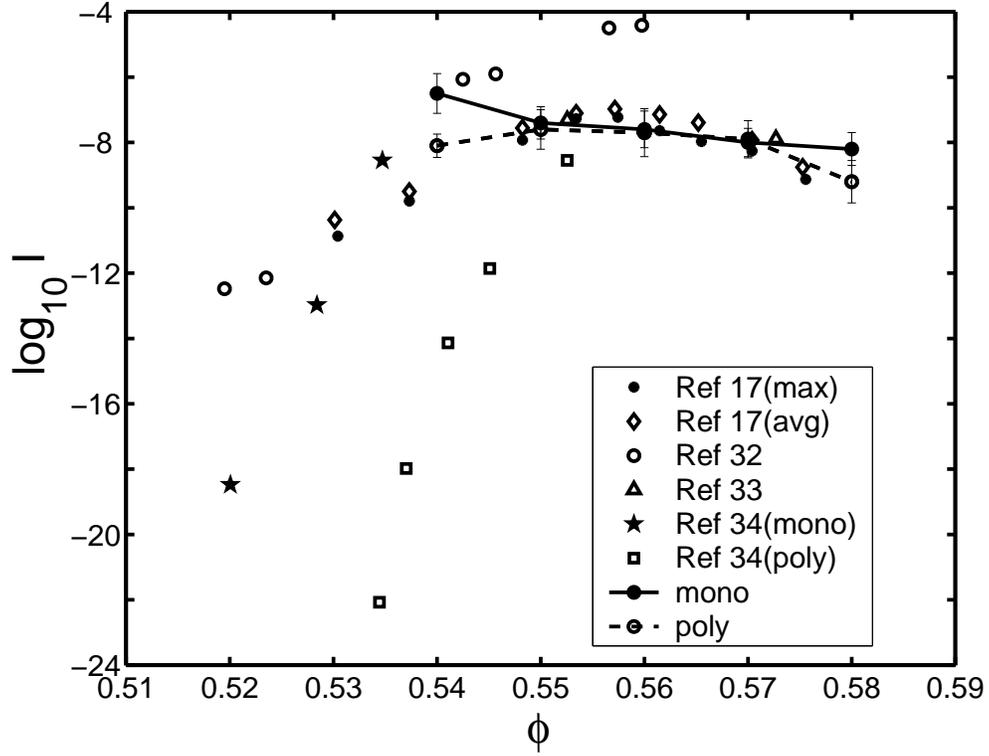}
\caption{\label{fig:cnr} The logarithm of crystal nucleation rates
(in units of $D_0/(2 R)^5$) for different packing fractions  for
the polydisperse and monodisperse systems (solid lines). The
experimental results as well as the results from other simulations
are shown for comparison. }
\end{figure}

By analyzing the time variations of the static structure factor we
were able to calculate the crystal fraction $X$, the average
linear crystal size $L$ and the number density of the crystals
$N_c$ (Fig.~\ref{fig:xlnc}). In spite of the minuscule systems
studied in the simulations, the time dependence is qualitatively
similar to the experimental data. Fig.~\ref{fig:cnr} shows a
summary of our results both for the polydisperse case (with $5 \%$
of polydispersity) and monodisperse systems together with the
experimental data \cite{17,19,21} and Monte Carlo simulations of
Auer and Frenkel \cite{7-1}. The latter simulations used the
umbrella sampling method in order to determine the probability of
the formation of the critical size nuclei and the free-energy
barrier for nucleation of a homogeneous crystal. This allowed them
to get the values of the crystal nucleation rates within the
framework of classical nucleation theory. Somewhat surprisingly,
their results were several orders of magnitude smaller than the
corresponding experimental results. In contrast, our results are
in a good agreement with the experimental data. The nucleation
rates for the polydisperse systems (especially for the lowest and
the highest concentrations studied) confirm the well established
fact that the presence of polydispersity slows the crystallization
down significantly.

However, due to the small size of the systems studied in the
simulations, such parameters as the average linear size and the
number density of the crystals cannot be determined directly. We
have found that, based on the structure factor analysis, the
average crystal size of the fully crystallized system is about
$0.5 - 0.8$ of the box size. On the other hand, the
local-invariant based calculation of the number of crystallites in
our systems indicates that there is only one crystallite at the
end of the crystallization process.

\begin{figure}
\includegraphics[scale = 0.7]{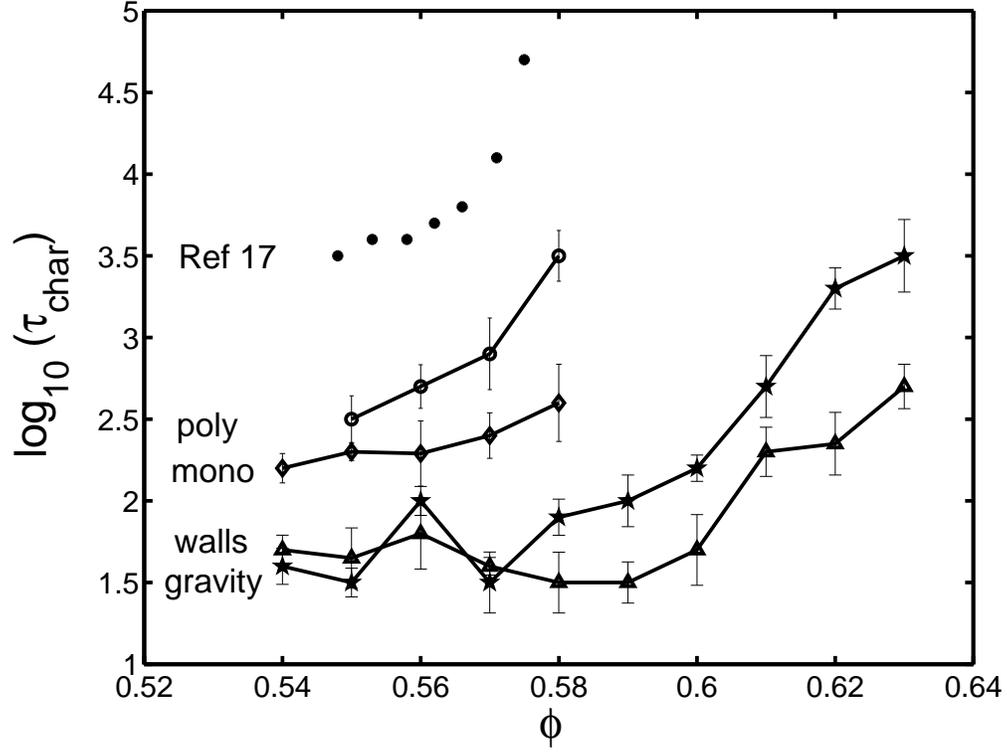}
\caption{\label{fig:times} The logarithm of the characteristic
crystallization times (measured by the crossover times - see text)
in units of the diffusional characteristic times for the following
cases: unbounded system(monodisperse and polydisperse cases), the
system with walls and the system in the presence of
gravity(monodisperse case). The experimental data from
Ref.~\cite{17} are shown. }
\end{figure}

Although we have found the crystal nucleation rates to be in a
good agreement with experimental results, the characteristic times
for the crystallization do not quite agree. The first difference
is the absence of an induction time\cite{17}, defined as the time
before the initialization of the crystallization. In all of the
systems studied here, the crystallization starts right after the
beginning of the simulation. The second difference is in the
values of the crossover times. The crossover time is defined as a
duration of crystallization that takes place at an approximately
uniform rate. Beyond the crossover time, the crystallization rate
slows down and is no longer constant. Our crossover times more
than $10$ times smaller than the corresponding experimental values
\cite{17} (see Fig.~\ref{fig:times}). To check whether this
discrepancy is an artifact of the small size of the system, we ran
a few simulations with $20000$ particles. The results were found
to be approximately the same indicating that the size dependence
is somewhat weak. Still, we observed the expected differences
between the polydisperse and monodisperse systems: the
crystallization processes were slower in the polydisperse systems.

\section{\label{sec:gravity}The effects of the bounding walls and the
gravity}

In order to investigate the dynamics of a system in the presence
of the gravitational field, it is essential to first bound the
system by some kind of walls. Otherwise we would deal with a free
fall situation when all of the processes proceed in exactly the
same way as in the absence of the gravity. Thus a good starting
point is to consider the system bounded in one dimension and
without any gravitational forces.

\begin{figure}
\includegraphics[scale = 0.7]{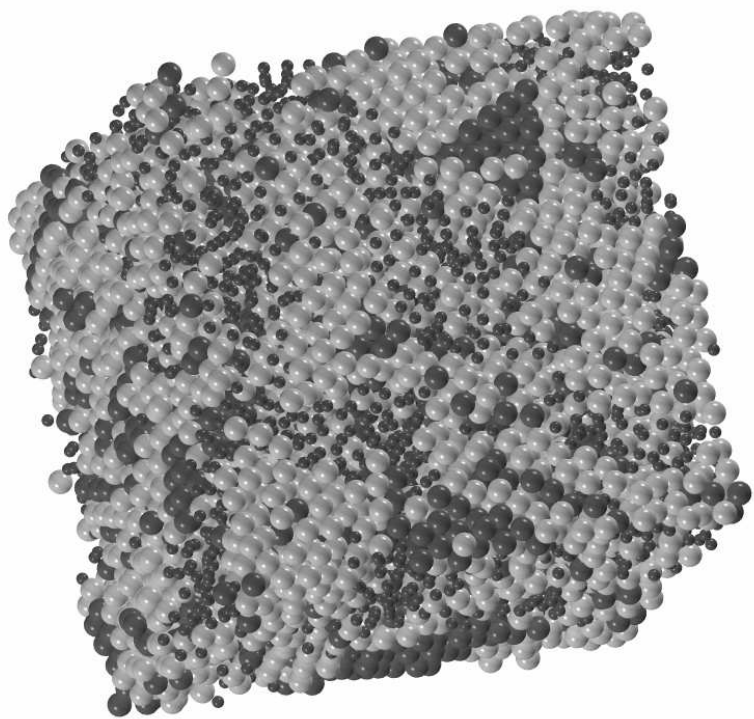}
\caption{\label{fig:sn} The snapshot of a system with volume
fraction of $\phi = 56\%$ and periodic boundary conditions in the
middle of the crystallization process. Here, small dark, large
light gray and large dark particles correspond to liquid, hcp and
fcc structures respectively. The liquid particle sizes have been
reduced to half their value for easier visualization. }
\end{figure}

\begin{figure}
\includegraphics[scale = 0.7]{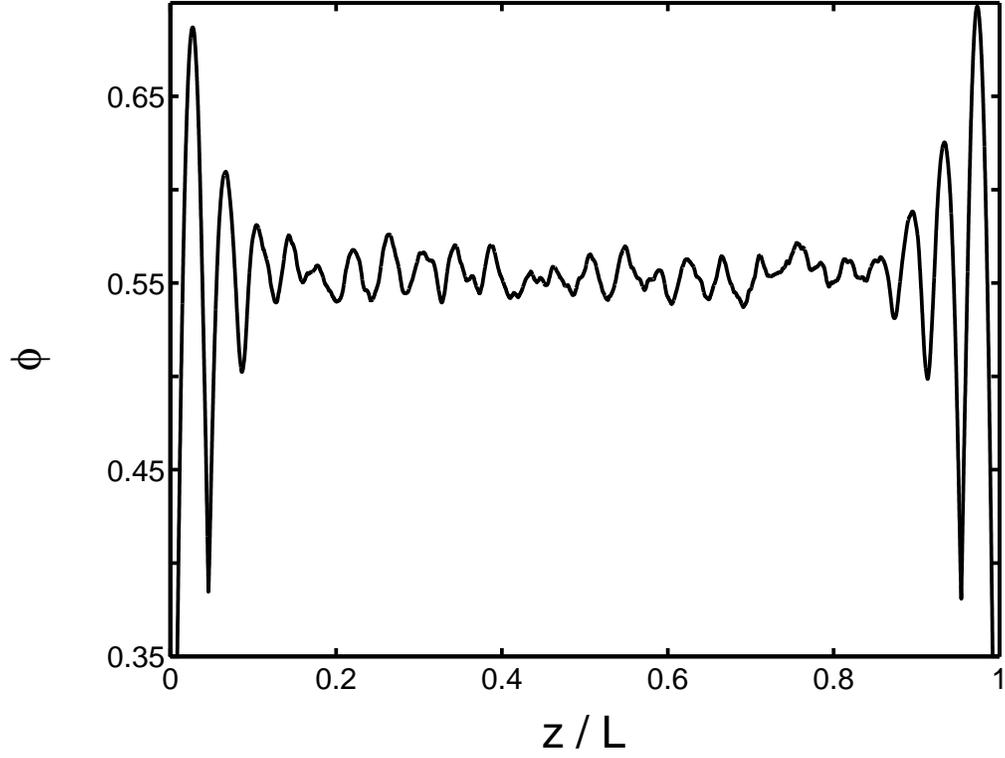}
\caption{\label{fig:sw0} The density profile of a typical initial
configuration of the system bounded by two walls with $\phi =
55\%$. $L$ is the distance between the two walls and is equal to
$21.86$ hard sphere diameters. }
\end{figure}

\begin{figure}
\includegraphics[scale = 0.7]{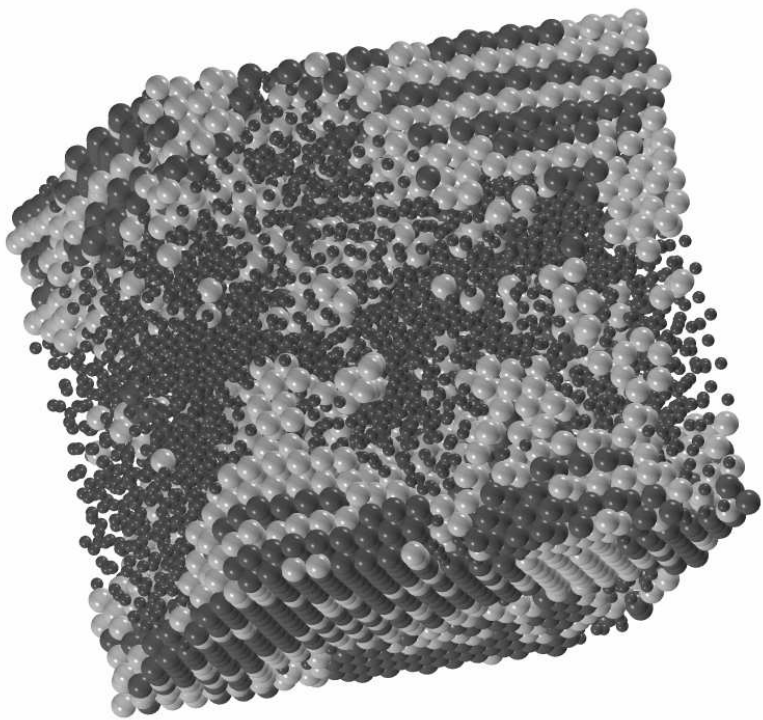}
\caption{\label{fig:sw} The snapshot of a system bounded by two
walls in the middle of the crystallization process. The convention
for the colors is as in Fig.~\ref{fig:sn} and again $\phi = 56\%$.
}
\end{figure}

The snapshot of the hard sphere configuration shown in
Fig.~\ref{fig:sn} indicates the complicated nature of
crystallization when periodic boundary conditions are used. Even
at moderately early stages of crystallization, there is an
interconnected phase of growing crystal nuclei with predominantly
hcp and fcc structures. The situation is significantly simpler
when walls are introduced. Even in the initial configuration (see
Figure~\ref{fig:sw0} for a typical example), there is pronounced
layering near the flat walls. These layers lead to a heterogeneous
wall-induced nucleation with growth of the crystal occurring
towards the center of the channel (Figure~\ref{fig:sw}).
Furthermore, the crystallization is more rapid compared to the
case without bounding walls, as seen in Fig.~\ref{fig:times}.

\begin{figure}
\includegraphics[scale = 0.7]{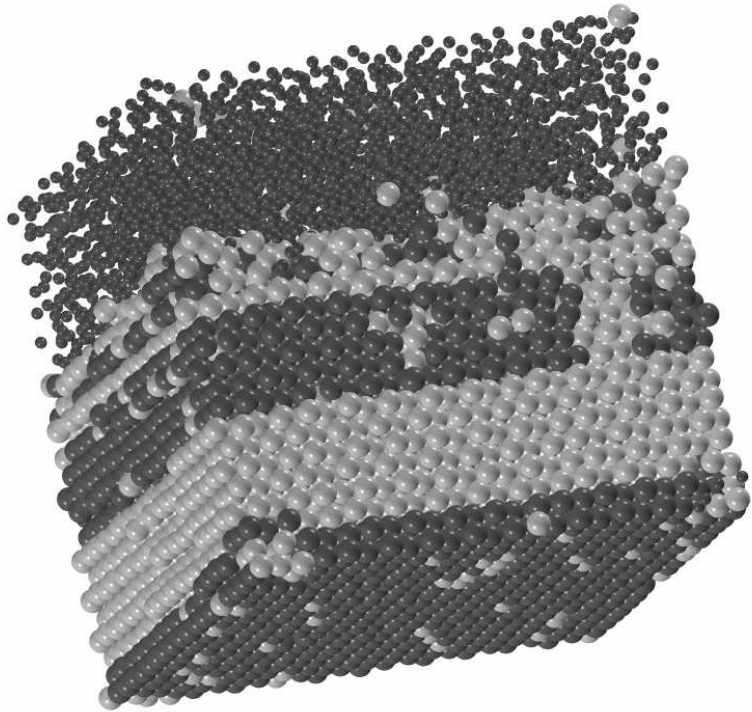}
\caption{\label{fig:swg1} The snapshot of a system ($\phi = 56\%$)
bounded by two walls in the presence of gravity (acting
downwards). The color code is as in Figure~\ref{fig:sn}. }
\end{figure}

\begin{figure}
\includegraphics[scale = 0.7]{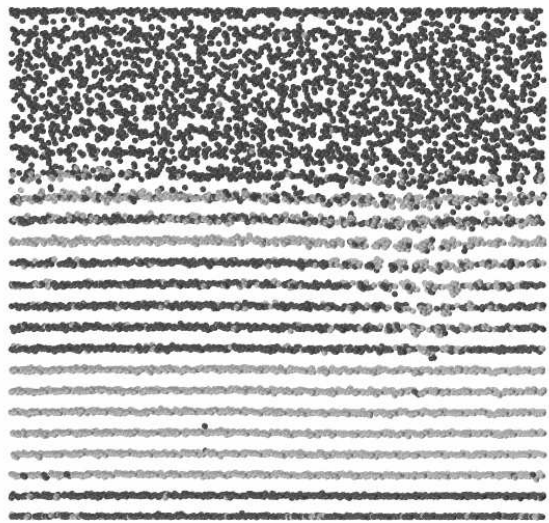}
\caption{\label{fig:swg2} The snapshot of the system shown in
Fig.~\ref{fig:swg1} rotated so that the crystal planes are
perpendicular to the image. The image sizes of the particles are
greatly reduced. Note the excellent planar ordering. }
\end{figure}

When the gravitational field in the direction perpendicular to the
bounding walls is turned on, the process of crystallization
switches to a different mode (see the snapshots shown in
Fig.~\ref{fig:swg1} and \ref{fig:swg2}). The particles are seen to
first settle down at the bottom of the channel, and after a while
we observe a stationary phase separation with the crystal at the
bottom and the liquid at the top of the channel. Note, that the
crystalline region consists of almost ideal hcp crystal planes
which are parallel to the bounding plane, whereas in the absence
of gravity, the crystallites are stacked at random orientations.

\begin{figure}
\includegraphics[scale = 0.7]{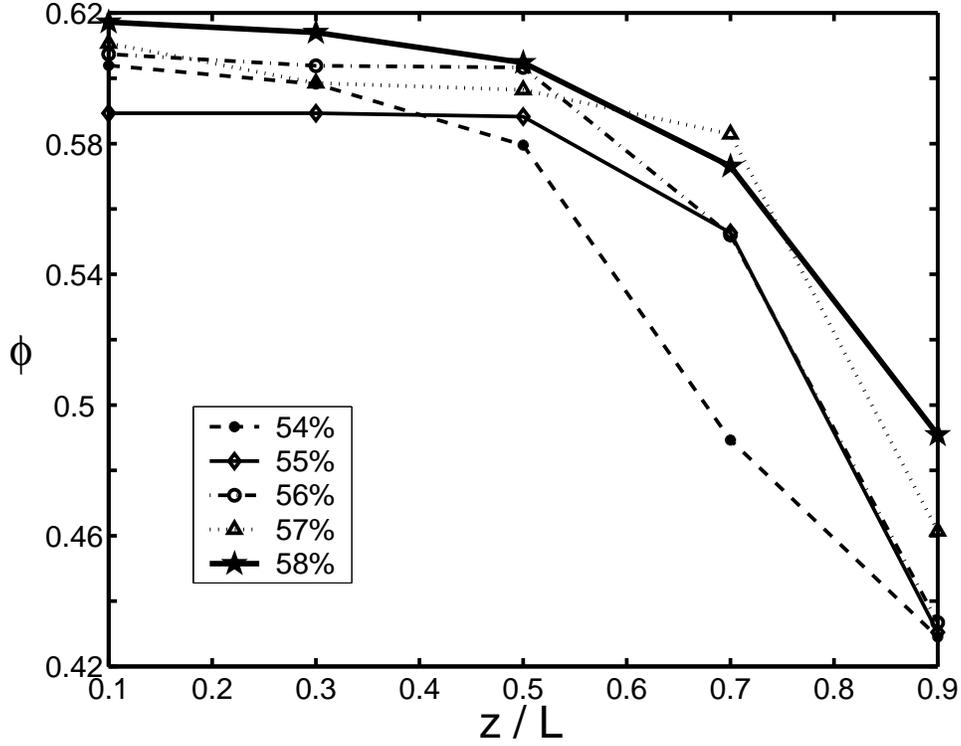}
\caption{\label{fig:cp} The concentration profile, as measured by
$\phi$, for the system in the presence of the gravitational field.
Here, $L$ is the distance between the two walls and is equal to
$\displaystyle{\left({\pi N\over 6 \phi_{total}}\right)^{1/3}}$
hard sphere diameters, where $N = 10976$ is the number of
particles and $\phi_{total}$ is the total concentration of the
systems (shown in the legend). The accelerations due to gravity
are $g_{54} \approx 4.55$, $g_{55}\approx 4.57$, $g_{56}\approx
4.60$, $g_{57}\approx 4.63$ and $g_{58}\approx 4.66$. }
\end{figure}

Fig.~\ref{fig:cp} shows the variation of the concentration with
the height, counting from the bottom plane. The concentration at
the bottom varies from $\phi \approx 58 \%$ to $\phi \approx 63
\%$ exceeding the average concentration by approximately
$3$-$5\%$. The concentration does not change significantly for up
to half of the channel and then it comes down to $\phi \approx 43
- 52 \%$ at the top where the system becomes a liquid.  We also
notice that the density profiles depend on the initial
concentrations only weakly, although the higher the initial
concentration of the system, the lower the propensity for the
group of particles to remain liquid-like. Interestingly, for the
system at $\phi = 58 \%$ the long time crystallization fraction is
about $98 \%$, although one can see from Fig.~\ref{fig:cp} that
about $20 \%$ of the volume of the system have concentrations
which are smaller than the melting value ($54.5 \%$). This can be
explained as emergence of ``induced'' crystallization, i.e.
crystallization promoted by the well-formed substrate \cite{lin1}.

While at concentrations up to $\phi \approx 58 \%$ the
crystallization times for the bounded systems with and without the
gravity are approximately the same, at higher concentrations we
observe that the presence of gravity slows the crystallization
down significantly. Thus, gravity stabilizes the glassy state by
reducing the mobility of the particles even though the presence of
the walls helps the crystallization. We observe crystallization in
the monodisperse systems at packing fractions as high as $63\%$,
which would lead to the glassy behavior in the absence of the
walls.

\section{\label{sec:structure}The crystal structure}

\begin{figure}
\includegraphics[scale = 0.7]{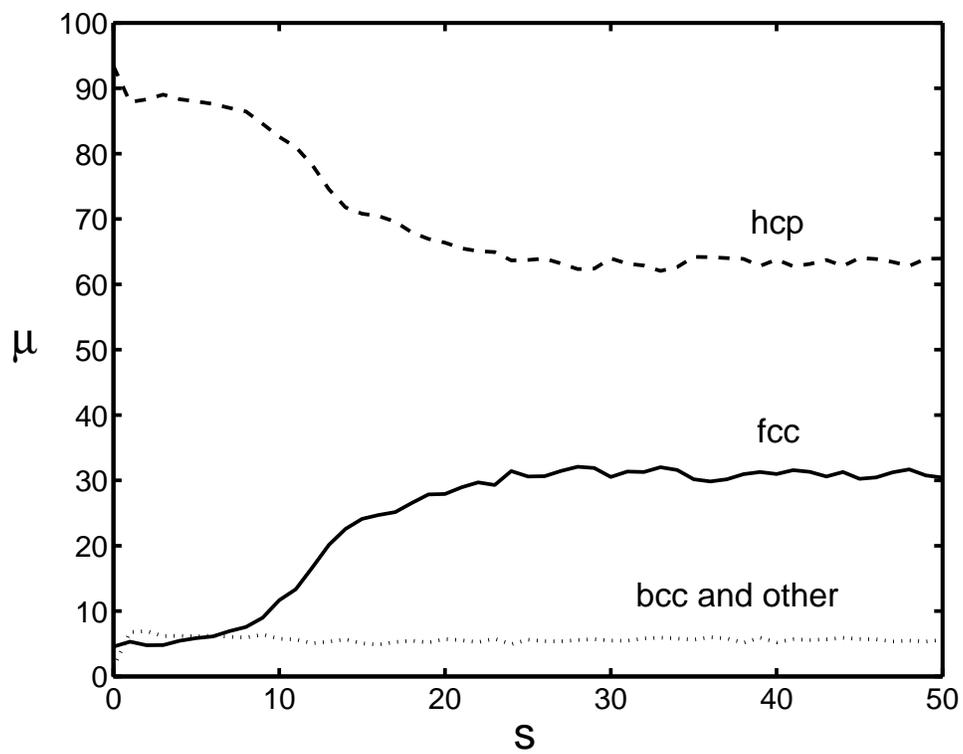}
\caption{\label{fig:frac} Plot of the fractions $\mu$ of fcc, hcp
and bcc structures in the system with $\phi = 56 \%$ versus $s$
defined as the number of collisions (in units of $1000\times N$,
where $N$ is the number of particles). }
\end{figure}

Finally, we have analyzed the nature of the crystal structure for
all of the cases studied here. The example of the structure for
one of the simulations ($\phi = 56 \%$, monodisperse, no walls,
zero gravity) is shown in Fig.~\ref{fig:frac} as a function of
time. The figure shows the percentage of the different crystal
types among the particles in the crystallized regions. The hcp
structure dominates in the first stages of crystallization. As the
crystallization proceeds, the fcc structure emerges and starts
growing.  In some cases, the fcc structure reaches a value equal
to $60 \%$ of the crystallized volume. The bcc (and other
packings) typically accounted for no larger than $5-10\%$ of the
number of crystal-like hard spheres. Once the crystallization is
completed, we do not observe any changes in the local structure.
Our observations allow us to conclude that the fcc structure is
more stable than the hcp especially because the fraction of the
fcc crystals
never decreases during the crystallization process \cite{gasser1,
huse1, woodcock1}.

\begin{acknowledgments}
This research was supported by the NASA Microgravity Fluids
Program and by KBN (grant number 2P03B).
\end{acknowledgments}

\bibliography{article_new}

\clearpage

\begin{table}
\caption{\label{tab:table1}The values of $q_4(i)$, $q_6(i)$ and
$\hat{w}_6(i)$ for different perfect crystal
structures\cite{16-1}. }
\begin{ruledtabular}
\begin{tabular}{cccc}
        & $q_4$ & $q_6$ & $\hat{w}_6$  \\
\hline
fcc     & 0.191 & 0.575 & -0.013 \\
hcp     & 0.097 & 0.485 & -0.012 \\
bcc     & 0.036 & 0.511 & 0.013  \\
sc      & 0.764 & 0.354 & 0.013  \\
icosahedral & 0 & 0.663 & -0.170 \\

\end{tabular}
\end{ruledtabular}
\end{table}

\clearpage

\clearpage

\clearpage

\clearpage

\clearpage

\clearpage

\clearpage

\clearpage

\clearpage

\clearpage

\clearpage

\clearpage

\clearpage

\end{document}